\documentclass{PoS}
\usepackage{units}
\usepackage{comment}
\usepackage{amsmath}

\title{Trigger developments for the fluorescence detector of EUSO-TA and EUSO-SPB2}

\ShortTitle{EUSO-TA/EUSO-SPB2 L1 trigger logic}

\author{
M.~Battisti$^{1}$, 
M.~Bertaina$^{1}$, 
F.~Fenu$^{1}$, 
H.~Miyamoto$^{1}$, 
\speaker{K.~Shinozaki}$^{1}$, 
A.~Belov$^2$ 
F.~Bisconti$^3$
M.~Mignone$^3$, 
F.~Capel$^4$

for the JEM-EUSO Collaboration \footnote{for the full JEM-EUSO Collaboration list see PoS(ICRC2019)1177} \\
\llap{$^1$} University of Turin and INFN Torino, Turin, Italy\\
\llap{$^2$} Moscow State University, Moscow,  Russia\\
\llap{$^3$} INFN Torino, Turin, Italy\\
\llap{$^4$} Royal Institute of Technology, Stockholm, Sweden\\
E-mail: \email{matteo.battisti@edu.unito.it}
}

\abstract{ The JEM-EUSO program aims at detecting Ultra High Energy Cosmic Rays (UHECRs) by observing the fluorescence light produced by extensive air showers (EAS) in the Earth's atmosphere. Within this program, a new generation of missions is being built, including (i) Mini-EUSO that will be installed on board the ISS in August 2019, (ii) an upgrade to the ground-based telescope EUSO-TA and (iii) the second super pressure balloon flight (EUSO-SPB2). All these detectors will have a dedicated trigger system based on a board equipped with a Xilinx Zynq device that will be able to detect different types of events on three different time-scales: a microsecond time-scale for cosmic ray detection (L1), a hundreds of microsecond time-scale for slower events like transient luminous events (TLEs) (L2), and a tens of millisecond time-scale used to produce a continuous monitoring, for even slower events like meteors or nuclearites. The L1 trigger logic for  the upgrade of EUSO-TA and EUSO-SPB2 are being developed taking into account the peculiarity of each detector (optic system, FOV, frame length) starting from the logic already developed for Mini-EUSO. In particular, every pixel will have an independent threshold that will be dynamically adapted to the level of the background; a predetermined condition on the number, the position and the time distribution of pixels above threshold has to be satisfied in order to issue a trigger. This contribution will summarize the L1 trigger logics and the tests currently performed.
}

\FullConference{36th International Cosmic Ray Conference -ICRC2019-\\
		July 24$^{th}$ - August 1$^{st}$, 2019\\
		Madison, WI, U.S.A.}

\begin{document}
\section{Introduction}
The JEM-EUSO project aims at detecting UHECRs from space through a fluorescence detector pointed towards the Earth's atmosphere \cite{EUSO-Program}. The main element that composes the focal surface of all EUSO-like detectors is the so-called Photo Detection Module (PDM), a square array of 36 Hamamatsu MAPMTs (Multi-Anode Photo Multiplier Tube), each of which is a 64-channel (8$\times$8 pixel) device capable of single photon counting (Fig. \ref{fig:EUSO-TA SPB and PDM}, Right). 4 MAPMTs are grouped in 2$\times$2 matrix called Elementary Cell (EC), therefore a PDM consists of 9 ECs, 2304 pixels in total \cite{PDM} (Fig.\ref{fig:EUSO-TA SPB and PDM},right). 

A trigger system for such a detector should be able to recognize a fluorescence signal lasting a few tens of $\mu s$, while keeping the trigger rate on the level of 1 Hz/PDM. A trigger logic optimized for the first long duration balloon flight (EUSO-SPB1) \cite{SPB1}, called PTT algorithm \cite{PTT}, was working grouping 9 pixels (3$\times$3 cell) together. The performance of the logic was tested  both in a field test and on flight, providing good results in terms of energy threshold and background rejection \cite{SPB1_trigger_performances}. Its main drawback was the threshold set at MAPMT level, since setting the same threshold for every pixel in a MAPMT could lead to an artificial and most important uncontrolled increment of the threshold for some pixels. Based on that idea, a new set of trigger logics is being developed, taking advantage of the logic already developed for Mini-EUSO \cite{Mini-EUSO_trigger}, where every pixel must have its own independent threshold, given the large footprint of each pixel on ground.

The data acquisition and trigger system of the new generation of detectors (Mini-EUSO \cite{Mini-EUSO}, the upgrade of EUSO-TA \cite{Lech_PoS} and EUSO-SPB2 \cite{SPB2}) is based on a board equipped with a Xilinx Zynq device that will allow to detect different events on three different time-scales. The microsecond time-scale and the hundreds of microsecond time-scale will have its own dedicated trigger system, called respectively L1 (intended for cosmic rays detection) and L2 (for TLEs and slower events), while a tens of millisecond time-scale is used to produce a continuous monitoring, for even slower events like meteors or nuclearites.

 Whenever an event satisfies the trigger logic, 128 temporal frames (GTU, Gate Time Unit) are stored in the board's internal memory and sent to the CPU. The Zynq board allows to store only a limited amount of L1 triggered events gated in 5.24s, (4 for Mini-EUSO, more than 4 for EUSO-TA), implying therefore the optimal trigger rate to be around or below 1 Hz. The task of the future L1 trigger logic will thus be to have a trigger rate as close as possible to 1 Hz, while having a suitable energy threshold.

\section{L1 Trigger logic general features}
\label{sec:trigger_general}
The general idea of EUSO-TA and EUSO-SPB2 L1 trigger logic is to have an adaptive threshold ($S\textunderscore pixel$) independent for each pixel, and then simply counts the number of active pixels in a certain portion of the PDM; an active pixel is defined as a pixel above its threshold.

The procedure to set an independent threshold for each pixel 
works as follow:
\begin{itemize}
	\item Every 128 GTUs the average value of each pixel is computed;
	\item The threshold ($S\textunderscore pixel$) for the next 128 GTUs is set $n_{\sigma}$ sigma above the average value, considering the background poissonian distributed. Every pixel has its independent threshold;
\end{itemize}

An adaptive threshold set to 4 $\sigma$ above the background (i.e. the value chosen both for the logic of EUSO-TA and EUSO-SPB2) guarantees by itself a very low probability of a pixel being active thanks to poissonian fluctuations of the background. For example, if the average value of a pixel is 1.5 counts/GTU, the threshold will be set to 7 counts/GTU. The probability to have at least 7 counts in a single pixel is thus $\sim 0.093 \%$. The different requests on the number, the position and the time distribution of active pixels, that will reduce the trigger rate by several orders of magnitude, depends on the characteristics of the detector, its optic system and the pointing direction and will be presented in the next sections.

\section{The EUSO-TA trigger logic} 
\begin{figure}[hbtp]\centering
\includegraphics[width=.5\textwidth]{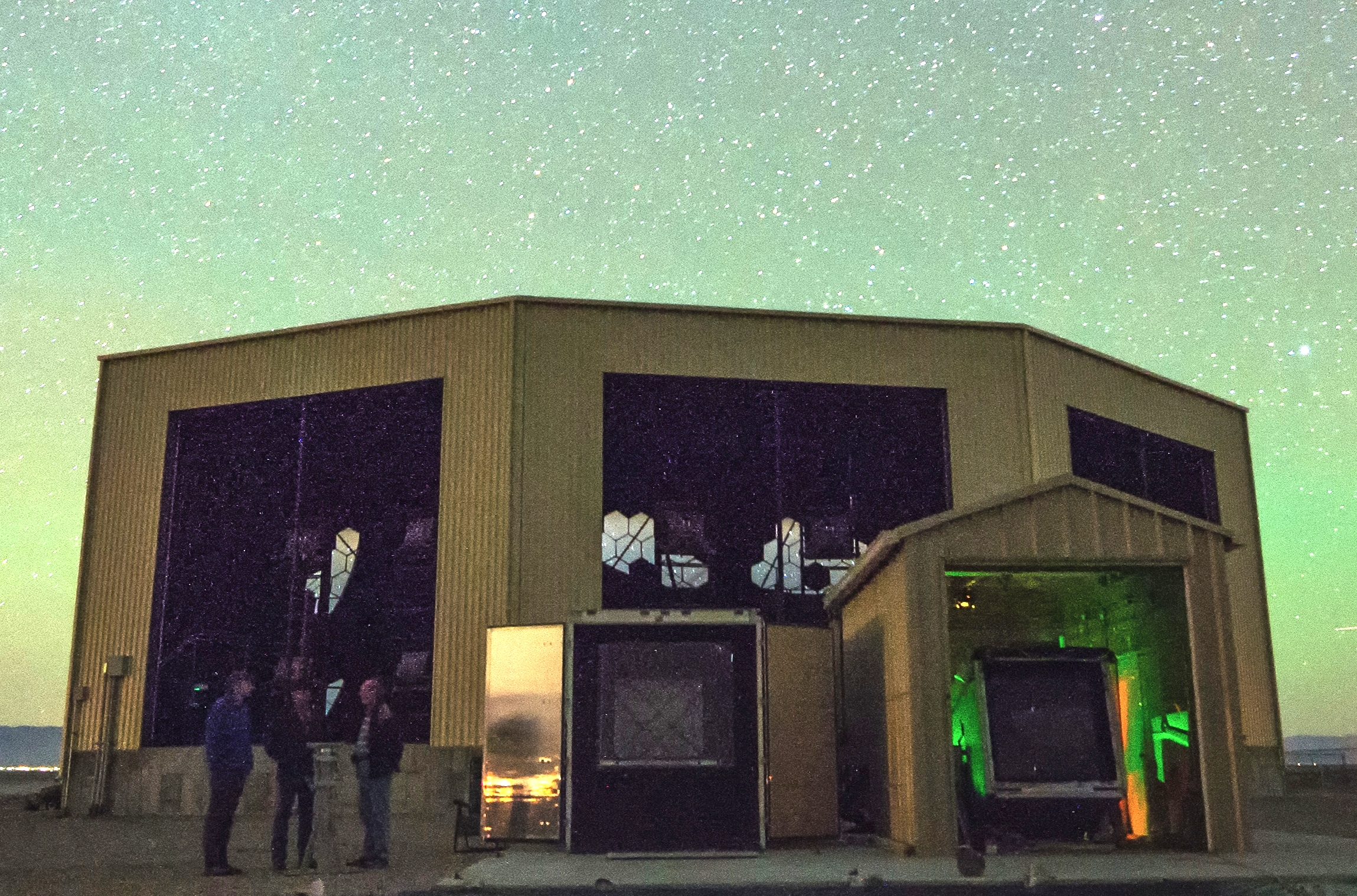} 
\includegraphics[width=.3\textwidth]{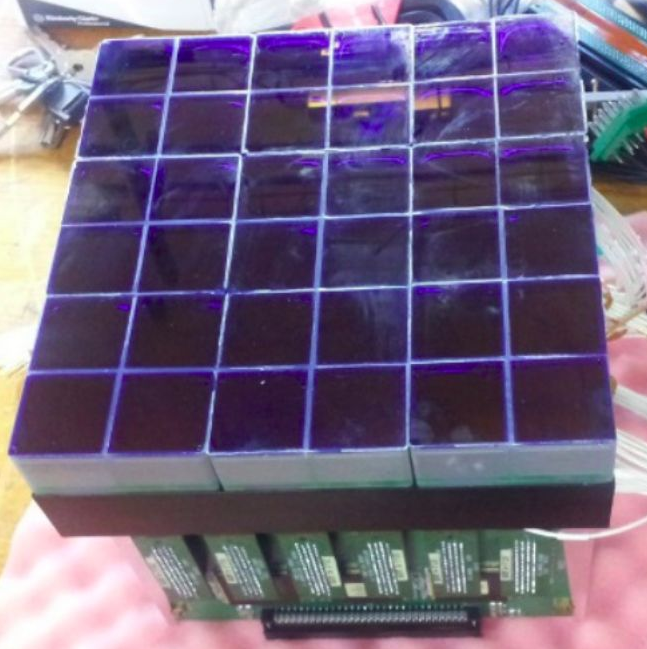} 
\caption{ Left: EUSO-TA telescope next to the EUSO-SPB1 PDM during cross-calibration procedure in 2016. EUSO-TA is the detector on the right, located in a shed next to a fluorescence detectors of the TA experiment. Right: a PDM (Photo Detection Module), EUSO detectors focal surface, made of 36 MAPMT, for a total of 2304 pixels }
\label{fig:EUSO-TA SPB and PDM}
\end{figure}

EUSO-TA is a ground-based telescope, installed at the Telescope Array (TA) site in Black Rock Mesa, Utah, USA in 2013. The telescope is housed in a shed located in front of one of the fluorescence detectors of the TA experiment (Fig.\ref{fig:EUSO-TA SPB and PDM},left).
\cite{TA_first_results}.

EUSO TA will go through an upgrade phase scheduled in 2020 in which many aspects of the detector will be improved. In particular, a Zynq board based DAQ will be installed, allowing an autonomous trigger system to be implemented. The temporal frame duration (GTU) will remain of 2.5 $\mu s$.

\subsection{EUSO-TA L1 Trigger logic}

Since EUSO-TA is ground-based, the distance of the closest approach between the EAS and the optic axis can vary, at least in principle, from several hundreds of meters to several kilometers. Therefore the signal can cross the entire PDM in just 1 GTU if the shower is close to the detector, or insist on the same MAPMT for several GTUs if the shower is farther away. That was taken into account setting different conditions for 1 GTU and 2 GTUs lasting signals.
Following the procedure explained in the previous section, the threshold is updated every 320$\mu s$ (128 GTUs) and set 4 $\sigma$ above the average value of each pixel. A trigger is issued whenever at least one of the following conditions is satisfied:
\begin{itemize}
			\item more than $n_{PMT_1}$ (i.e. 5) pixels above threshold in the same MAPMT in a single GTU;
			\item more than $n_{PMT_2}$ (i.e. 6) pixels above threshold in the same MAPMT integrating over 2 consecutive GTUs;
			\item more than $n_{EC_1}$ (i.e. 7) pixels above threshold in the same EC in a single GTU;
			\item more than $n_{EC_2}$ (i.e. 9) pixels above threshold in the same EC integrating over 2 consecutive GTUs;
			\item more than $n_{PDM_1}$ (i.e. 15) pixels above threshold in the entire PDM in a single GTU;
			\item more than $n_{PDM_2}$ (i.e. 20) pixels above threshold in the entire PDM integrating over 2 consecutive GTUs.
\end{itemize}

A low energy event close to the detector will produce a fast and faint signal on a huge portion of the PDM, therefore the threshold $n_{EC_1}$ or $n_{PDM_1}$ are more likely to be responsible for the triggers. On the other hand, a more distant shower will produce a longer track that will insist on the same MAPMT for a few GTUs, and will be hopefully detected by the thresholds $n_{PMT_1}$ or $n_{PMT_2}$.

Values in brackets, which represent the best trade-off between a high rejection power and high sensitivity, were set after tests performed over data acquired during the previous campaign of EUSO-TA and EUSO-SPB1 \cite{SPB1}. If needed these values can be easily modified.

\begin{figure}[hbtp]\centering
\includegraphics[width=.96\textwidth]{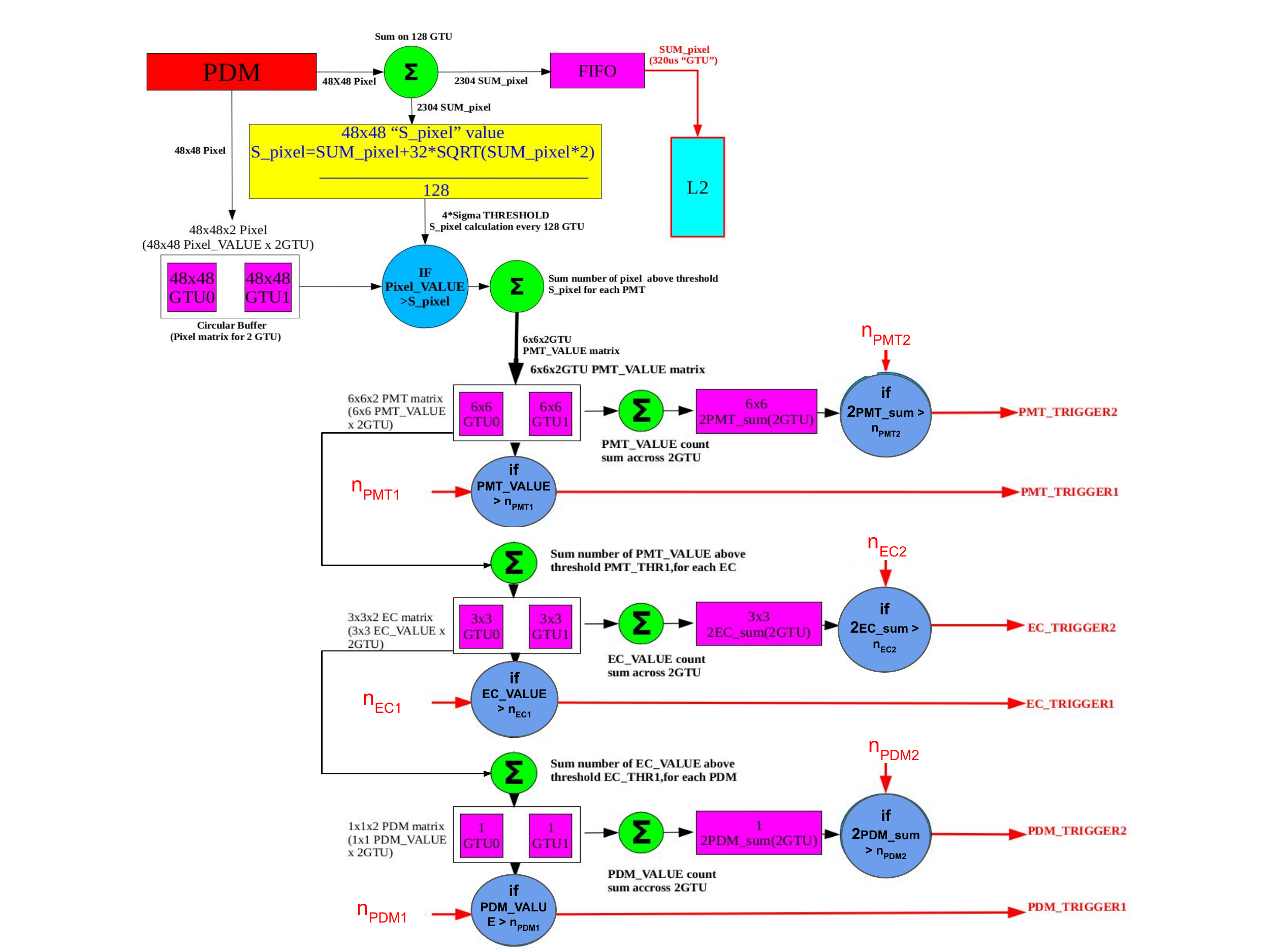}
\caption{ Scheme of firmware implementation of EUSO-TA trigger logic.}
\label{fig:VHDL_EUSO-TA}
\end{figure}

The expression of S\textunderscore pixel in Fig. \ref{fig:VHDL_EUSO-TA} is the formula used by the FPGA to compute the 4 sigma threshold. Calling ${SUM}$ the value of the integration of the counts of a pixel over 128 GTUs, the average background value $\lambda$ will simply be $\lambda = \frac{{SUM}}{128}$. The threshold set to 4 $\sigma$ will then be:

\begin{equation} 
\begin{split}
{S\textunderscore pixel}& = \lambda + n_{sigma} \cdot \sqrt{\lambda} = \frac{{SUM}}{128} + 4 \cdot \sqrt{\frac{{SUM}}{128}}  =\frac{{SUM} + 32  \cdot \sqrt{{SUM} \cdot 2}}{128} 
\end{split}
\end{equation}

The probability to have a trigger at MAPMT level is the probability of having at least 6 pixels above threshold in a MAPMT in a GTU, or 7 in two consecutive GTUs, and it is given by the cumulative distribution function of the binomial distribution (keeping in mind that in a MAPMT there are 64 pixels). Following the example of the previous section, given an average background value of 1.5 counts/GTU, this value is $\sim 5 \cdot 10^{-11}$ for both the conditions. In a PDM there are 36 MAPMTs, therefore the total probability of the trigger due to poissonian oscillation at MAPMT level is around $3.4 \cdot 10^{-9}$%RIGUARDA I CONTI!!!!
. Since in a second there are $4 \cdot 10^{5}$ GTUs, the fake trigger rate in this condition should be around $ 1.4 \cdot 10^{-3} $ $Hz$. Similar calculation can be performed for the threshold at EC and PDM level, obtaining trigger rate on the same order of magnitude. The total trigger rate is $\sim 5 \cdot 10^{-3}$ Hz. The results are even better if the average background is higher, a little worse with lower background. The biggest source of fake trigger may therefore come from non-poissonian events, like
electrical noise.

\subsubsection{Expected trigger rate}
EUSO-TA L1 trigger logic has been applied off-line to data taken in the previous EUSO missions. The expected trigger rate was estimated analysing a subset of EUSO-SPB1 flight data, obtained by removing the part with the triggered signal, for a total of almost 1 million of 2.5$\mu s$ GTUs.

\begin{figure}[hbtp]\centering
\includegraphics[width=.3\textwidth]{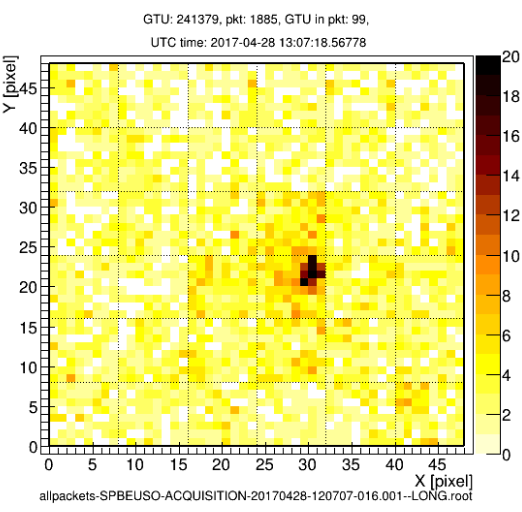}
\includegraphics[width=.3\textwidth]{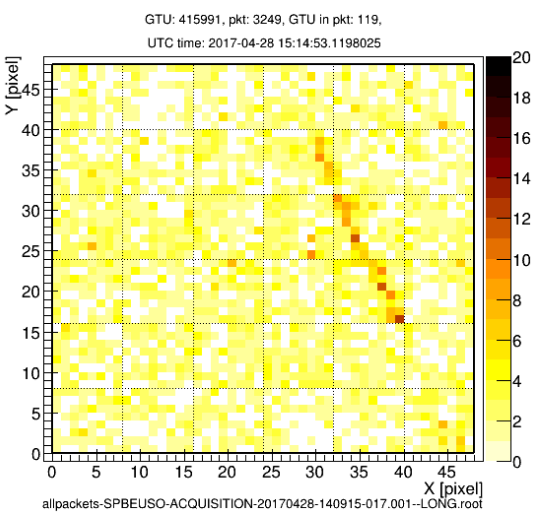}  
\includegraphics[width=.3\textwidth]{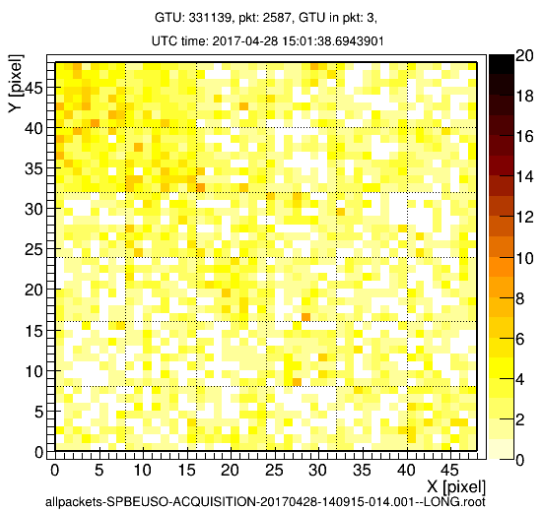}
\caption{ Examples of triggered events by EUSO-TA logic. Most of the trigger were caused by electrical noise or by low energy cosmic rays impinging on the focal surface. Left: example of small luminous blob. Center: track issued by direct cosmic ray. Right: electrical noise on one EC. }
\label{fig:Fake_triggers}
\end{figure}

Obviously the number of triggered events depends dramatically on the values of $n_{PMT_1}$ and the other thresholds. The chosen values, in brackets in the previous chapter, detected 10 triggers, for a trigger rate of 4.11 Hz. However most of the triggers are due to electrical noise or low energy cosmic rays directly hitting the focal surface (cfr. Fig.\ref{fig:Fake_triggers}), the "real" fake trigger rate should therefore be $\sim$ 1 Hz, which is close enough to the optimal trigger rate for EUSO-TA.

\subsubsection{Expected performances}
\begin{figure}[hbtp]\centering
\includegraphics[width=.75\textwidth]{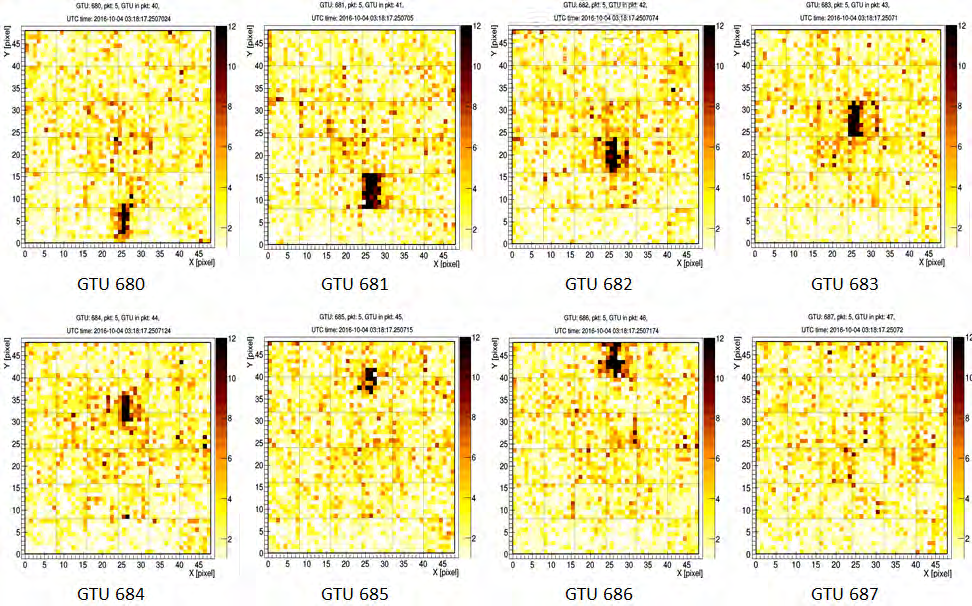}
\caption{ Laser event recorded during SPB1 Utah campaign }
\label{fig:SPBatUtah_event}
\end{figure}

The trigger efficiency has been estimated analysing the dataset of the field test at the Telescope Array site \cite{SPB1}, when the PDM of EUSO-SPB1 was installed next to EUSO-TA (Fig. \ref{fig:EUSO-TA SPB and PDM}, left). The dataset consists of 100 laser shots fired at a fixed energy, with a frequency of 4 Hz. The laser was fired at a distance of 22 km from the detector, the energy ranging from 3 mJ to 0.7 mJ. For each laser shot, if the SPB1 on-line trigger algorithm (PTT trigger \cite{PTT}) detected an event, one packet of 128 GTUs was stored. For the highest energy 100 shots are thus available, while this is not the case for lower energies when not all laser shots were triggered. An example of laser shot is reported in Fig.\ref{fig:SPBatUtah_event}

\begin{figure}[!hbtp]\centering
\includegraphics[width=.49\textwidth]{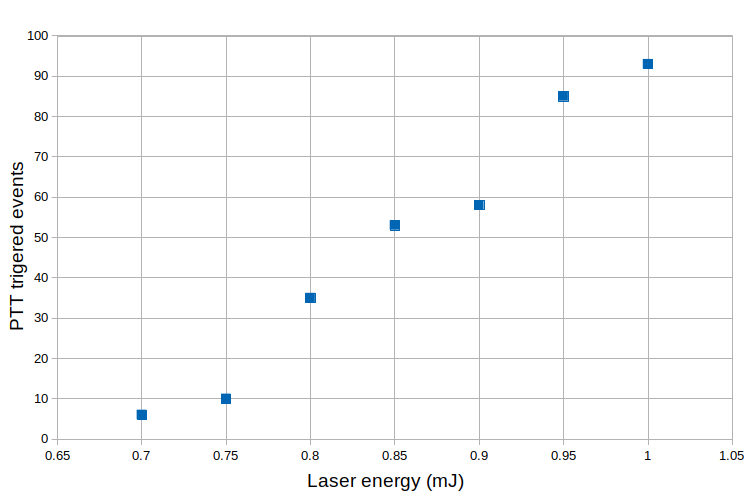}
\includegraphics[width=.50\textwidth]{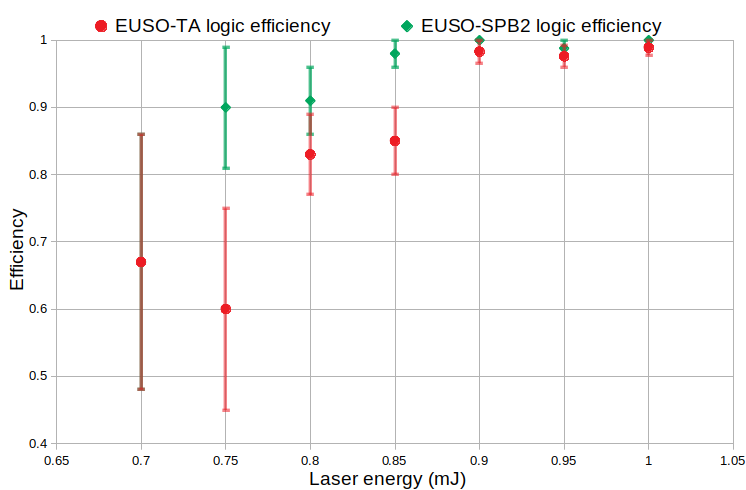}
\caption{ Left: number of events detected by the on-line PTT logic as a function of laser energy. Above 1 mJ the number of detected events is constantly between 97 and 100. Left: estimated efficiencies of EUSO-TA and EUSO-SPB2 trigger logic. Efficiency is defined as the ratio between the number of events detected by the off-line logic and the number of available events (i.e. the events detected by the on-line PTT logic).
Data are shown for laser energy below 1 mJ. Above this value, the efficiency of the logics is always compatible with 1.}
\label{fig:Performances}
\end{figure}

Once again the dataset has been analysed off-line, to estimate the expected performances. The plots in Fig. \ref{fig:Performances} show the number of triggered laser shots detected by the on-line logic (on the left) and the efficiency of the EUSO-TA trigger logic, computed analysing the triggered data off-line. Efficiency is here defined as the ratio between the number of events detected by the off-line logic and the number of available events (i.e. the events detected by the on-line PTT logic). In the plot is shown also the result of the logic foreseen for EUSO-SPB2 (also off-line, see next chapter).

For laser energy above 0.9 mJ the efficiency is constantly compatible with 1 since the signals are very bright and easy to detect. At lower energies the efficiency is always around or above 0.5. It is worth noting that, unlike PTT and SPB2 algorithms, EUSO-TA trigger logic is not optimized for distant events like the laser shots (22 km) and still its result is not significantly worse than the other logics.

\section{The EUSO-SPB2 trigger logic}
EUSO-SPB2 will be the second super-pressurised balloon flight \cite{SPB2}, built upon the experience of flying EUSO-SPB1 in the Spring of 2017. It is planned to fly in 2022 from Wanaka, New Zealand. Its main scientific purpose will be the detection of UHECRs with a fluorescence telescope based on three PDMs and the hunt for cosmogenic Earth-skimming tau neutrino with a Cherenkov telescope based on SiPMs \cite{SPB2}. One first level GTU in SPB2 will be 1$\mu s$ long, reducing therefore the duration of the second and third level GTUs by a factor of 2.5 . 

\subsection{EUSO-SPB2 L1 Trigger logic for the fluorescence telescope}
From a trigger point of view the three PDMs (and, as a matter of fact, each MAPMT) will be independent, when a trigger occurs the entire focal surface will be read out. 

As explained in section \ref{sec:trigger_general}, the threshold is updated every 128$\mu s$ (128 GTUs) and set 4 $\sigma$ above the average value of each pixel. A trigger is issued when both of the following conditions are satisfied:

\begin{itemize}
	\item more than $n_{pixel}$ (2) pixels above threshold in the same MAPMT in a single GTU;
	\item the same MAPMT active for at least $N_{GTU}$ (2) consecutive GTUs. A MAPMT is considered active when there are more than $n_{pixel}$  above threshold in a single GTU.
		
\end{itemize}

The request of a persistent signal in one MAPMT comes from the fact that the residence time of an EAS in a MAPMT FOV is of the order of a few $\mu s$. The values in brackets represent the best threshold to optimize efficiency and fake trigger rate. 

\subsubsection{Expected trigger rate and performances}
Trigger rate and performances of SPB2 trigger has been estimated in the same way done for EUSO-TA trigger logic. Once again the results depend critically on the values of $n_{pixel}$ and $N_{GTU}$: for the values in brackets above, it leads to an expected trigger rate between 2 and 4 Hz, and an efficiency at the same level or better than the one of PTT and TA logic (Fig. \ref{fig:Performances}).

\section{Summary}
The EUSO-TA trigger logic has been tested analysing data from previous EUSO missions. The results in terms of energy threshold and rejection power look quite promising, and will have to be confirmed with experimental measurements at TurLab \cite{TurLab} and in open sky conditions.

The logic for EUSO-SPB2 has been defined and tested analysing data from the previous balloon flight, proving itself capable of detecting laser-induced events while keeping the trigger rate to an acceptable level.
Once experimental data and Monte Carlo simulations with 1 $\mu s$ GTUs will be available, a better fine tuning of the threshold will be performed, if needed.
\section*{Acknowledgments}
%\noindent{\textbf{Acknowledgments}\\
{\footnotesize
This work was partially supported by Basic Science Interdisciplinary Research
Projects of RIKEN and JSPS KAKENHI Grant (JP17H02905, JP16H02426 and
JP16H16737), by the Italian Ministry of Foreign Affairs and International
Cooperation, by the Italian Space Agency through the ASI INFN agreement
n. 2017-8-H.0 and contract n. 2016-1-U.0, by NASA award 11-APRA-0058 in
the USA, by the
Deutsches Zentrum f\"ur Luft- und Raumfahrt, by the French space agency
CNES, the Helmholtz Alliance for Astroparticle Physics funded by the
Initiative and Networking Fund of the Helmholtz Association (Germany), by
Slovak Academy of Sciences MVTS JEMEUSO as well as VEGA grant agency project
2/0132/17,
by National Science Centre in Poland grant (2015/19/N/ST9/03708 and
2017/27/B/ST9/02162), by Mexican funding agencies PAPIIT-UNAM, CONACyT and
the
Mexican Space Agency (AEM). Russia is supported by ROSCOSMOS and the Russian
Foundation for
Basic Research Grant No 16-29-13065. Sweden is funded by the Olle Engkvist
Byggm\"astare Foundation.

 The support received by the Telescope Array Collaboration is deeply acknowledged.
}

\end{document}